\documentclass[sn-mathphys-num]{sn-jnl}

\usepackage{graphicx}
\usepackage{subfigure}
\usepackage{multirow}
\usepackage{amsmath,amssymb,amsfonts}
\usepackage{amsthm}
\usepackage{mathrsfs}
\usepackage[title]{appendix}
\usepackage{xcolor}
\usepackage{textcomp}
\usepackage{manyfoot}
\usepackage{booktabs}
\usepackage{algorithm}
\usepackage{algorithmicx}
\usepackage{algpseudocode}
\usepackage{listings}

\theoremstyle{thmstyleone}

\theoremstyle{thmstyletwo}

\theoremstyle{thmstylethree}

\begin{document}

\title[Article Title]{Thermal chaos of quantum-corrected-AdS black hole in the extended phase space}

\author[]{\fnm{Lei} \sur{You}}

\author[]{\fnm{Rui-Bo} \sur{Wang}}

\author[]{\fnm{Yu-Cheng} \sur{Tang}}

\author[]{\fnm{Jian-Bo} \sur{Deng}}

\author*[ ]{\fnm{Xian-Ru} \sur{Hu}}\email{huxianru@lzu.edu.cn}

\affil[]{\orgdiv{Lanzhou Center for Theoretical Physics, Key Laboratory of Theoretical
		Physics of Gansu Province}, \orgname{Lanzhou University}, \city{Lanzhou}, \postcode{73000}, \state{Gansu}, \country{China}}

\abstract{We briefly analyzed the equation of state and critical points of the quantum-corrected Schwarzschild-like black hole and used the Melnikov method to study its thermal chaotic behavior in the extended phase space of flat, closed, and open universes. The results show that the black hole's thermodynamic behavior is similar to that of the Van der Waals system. Although the critical ratios at the critical points differ among the three universes, they are all independent of the quantum correction parameter. For chaos, time perturbations will lead to chaotic behavior when their amplitude exceeds a critical value that depends on the quantum correction parameter and the radius of the dust sphere in the FRW model. Based on this, we found that the chaotic behavior of the black hole varies across different universes depending on the quantum correction parameter, but this parameter always makes chaos more likely. Using the value of the quantum correction parameter determined by Meissner, chaos is always more difficult to occur in an open universe compared to the other two types of universes. Which universe is most prone to chaos depends on the radius of the dust sphere.
Finally, chaotic behavior is always present under spatial perturbations.}


\maketitle

\section{Introduction}\label{sec1}

In the 1970s, Bekenstein and Hawking published seminal papers on black hole thermodynamics, skillfully correlating black holes with classical thermodynamic theory~\cite{bekens,hawking}. Since then, significant progress has been made, demonstrating that black holes exhibit phase structures and transitions similar to those of ordinary thermodynamic systems. Physicists have studied these phenomena in various contexts, particularly in anti-de Sitter (AdS) spacetime, which has a negative cosmological constant. Kubiznak et al. proposed an extended phase thermodynamics framework by treating the cosmological constant as a dynamical thermodynamic variable (pressure) and its conjugate quantity as thermodynamic volume \cite{kubizvnak2012p,kastor2009enthalpy}. Within this framework, studies of the $Pv$ critical behavior of Reissner-Nordström-AdS black holes and other AdS black hole systems have established a correspondence between small/large black hole phase transitions in the extended phase space and the liquid/gas phase transitions in the Van der Waals system \cite{kubizvnak2012p,dx1,dx2,dx3,dx4,dx5}. For the Van der Waals system, researchers like Slemrod et al. systematically studied its thermodynamic chaotic behavior using the Melnikov method~\cite{slemrod1985temporal,melnikov1963stability}. Consequently, there is a natural interest in exploring the thermal chaotic behavior of black holes in the extended phase space.

Chaos, a process of system evolution highly sensitive to initial conditions, is prevalent in nonlinear and non-integrable dynamical systems, including black hole physics~\cite{c1,c2}. Researchers have explored various chaotic phenomena in black holes. For example, Cornish and Levin identified chaotic orbits in close binary black hole systems \cite{cornish2003lyapunov}; Suzuki and Maeda observed that particle spin effects near a Schwarzschild black hole induce chaotic trajectories \cite{suzuki1997chaos}; and Kitaev's Sachdev-Ye-Kitaev (SYK) model highlighted strong chaotic behavior in simplified quantum gravity models \cite{kitaev2015simple}.

As previously mentioned, due to the similarity between black hole critical behavior and the Van der Waals system, the thermal chaotic behavior of black holes has been extensively studied in recent years \cite{rhd1,rhd2,rhd3,rhd4,rhd5}. Scholars, drawing an analogy with the chaotic behavior in the Van der Waals system, have introduced time and spatial periodic perturbations in the $Pv$ thermodynamic phase space of black holes and detected chaos by studying the zeros of the Melnikov function. For example, Chabab et al. found that Reissner-Nordström-AdS black holes might exhibit chaotic behavior under thermal perturbations, but this chaotic behavior disappears when the charge exceeds a certain threshold \cite{rhd1}. Mahish et al. discovered that under time perturbations, the presence of charge is necessary for the onset of chaos in Gauss-Bonnet-AdS black holes, whereas chaotic behavior always occurs under spatial perturbations, regardless of the charge \cite{rhd2}.

In this paper, we will study the thermal chaotic behavior of a recently proposed quantum-corrected black hole by Lewandowski et al. \cite{lewandowski2023quantum}. Zhang et al. investigated the additional halo caused by this black hole, providing new insights into verifying the quantum correction model \cite{zhang2023black,you2024decoding}. However, there has been little research on the thermodynamic properties of this black hole in the extended phase space, and how quantum effects influence its thermal chaos remains unknown. Particularly, this black hole retains the cosmological curvature $k$ from the FRW model, making it highly significant to study its thermal chaotic behavior in universes with different $k$ values. By studying these aspects, we aim to further enrich our understanding of chaotic behavior in black holes and investigate the effects of quantum gravity in different cosmological backgrounds.

The organization of this paper is as follows. In Sect.~\ref{sec2}, we briefly review the derivation of the quantum-corrected black hole solution and extend it to AdS spacetime. In Sect.~\ref{sec3}, we derive the equation of state for this black hole and discuss the influence of black hole parameters on the critical points. In Sect.~\ref{sec4}, we investigate the chaotic behavior of this black hole under periodic perturbations in time and space and explore the impact of black hole parameters on chaos. Finally, in Sect.~\ref{sec5}, we provide a summary and discussion.

\section{Quantum-corrected-AdS black hole solution}\label{sec2}

The spacetime manifold $\mathcal{M}$ is divided by the hypersurface $\Sigma=\mathcal{M}^{-}\cap\mathcal{M}^{+}$ into an interior manifold $\mathcal{M}^{-}$ and an exterior manifold $\mathcal{M}^{+}$. The interior manifold $\mathcal{M}^{-}$ is filled with a uniformly spherically symmetric distribution of irrotational dust, and its line element is described by the FRW metric as
\begin{equation}
	\mathrm{d}s_{-}^{2}=-\mathrm{d}\tau^{2}+a(\tau)^{2}\left(\frac{\mathrm{d}\tilde{r}^{2}}{1-k\tilde{r}^{2}}+\tilde{r}^{2}\mathrm{d}\Omega^{2}\right),
\end{equation} 
where $\tau$ is the proper time of the comoving observer, $a$ is the scale factor, $k$ determines the geometry of the universe (0 for flat, 1 for closed, or -1 for open), and $\mathrm{d}\Omega^{2}\equiv\mathrm{d}\theta^{2}+\sin^{2}\theta\mathrm{d}\phi^{2}$ is the spherical line element.

In the classical case, according to Birkhoff's theorem, the exterior vacuum solution for a spherically symmetric gravitational field is the Schwarzschild metric. To consider a more general case, we assume that the exterior manifold $\mathcal{M}^{+}$ is a static, spherically symmetric spacetime, with its line element given by
\begin{equation}
	\mathrm ds_{+}^{2}=-f(r)\mathrm dt^{2}+f(r)^{-1}\mathrm dr^{2}+r^{2}\mathrm d\Omega^{2}.
\end{equation}

Let the hypersurface $\Sigma$ be located at $\tilde{r}=\tilde{r}_{0}$. Applying the Darmois-Israel junction conditions at $\Sigma$ to match the interior and exterior spacetimes $\mathcal{M}^{-}$ and $\mathcal{M}^{+}$ yields~\cite{yang2023shadow}
\begin{equation}\label{af}
	\begin{cases}
		a(\tau)\tilde{r}_{0} =r(\tau), \\
		\\
		f(r) =\left(1-k\tilde{r}_{0}^{2}\right)-H(\tau)^{2}r^{2},
	\end{cases} 
\end{equation}
where $H(\tau)\equiv\frac{1}{a(\tau)}\frac{\mathrm{d}a(\tau)}{\mathrm{d}\tau}$ is the Hubble parameter. For the classical case, $H^2=\frac{8\pi G}{3}\rho-\frac{k}{a^2}$, substituting into Eq.~(\ref{af}) yields the Schwarzschild metric. For the effective quantum correction case, the effective Friedmann equation reads
\begin{equation}\label{heff}
	H_{\mathrm{eff}}^{2}=\frac{8\pi G}{3}\left(\rho-\frac{3}{8\pi G}\frac{k}{a^{2}}\right)\left(1-\frac{\rho-\frac{3}{8\pi G}\frac{k}{a^{2}}}{\rho_{c}}\right),\quad\rho=\frac{M}{\frac{4}{3}\pi\tilde{r}_{0}^{3}a^{3}},
\end{equation}
where the deformation parameter is the critical density $\rho_{c}=\sqrt{3}/(32\pi^{2}\gamma^{3}G^{2}\hbar)$ with the Barbero-Immirzi parameter $\gamma$ (determined to be $0.2375$ in \cite{ga1,ga2}), $M$ is the mass of the ball of dust with radius $a(\tau)\tilde{r}_0$ in the interior manifold $\mathcal{M}^{-}$ spacetime, and $G$ is the Newtonian gravitational constant. Substituting Eq.~(\ref{heff}) into Eq.~(\ref{af}) yields
\begin{equation}\label{fr}
	\begin{aligned}
		f(r)& =\left(1-k\tilde{r}_{0}^{2}\right)+\frac{8\pi G}{3}\left(\rho-\frac{3}{8\pi G}\frac{k}{a^{2}}\right)\left(1-\frac{\rho-\frac{3}{8\pi G}\frac{k}{a^{2}}}{\rho_{c}}\right)r^{2}  \\
		&=1-\frac{R_s}{r}+\frac{4\sqrt{3}\pi\gamma^3G\hbar}{r^2}\left(\frac{R_s}{r}-k\tilde{r}_0^2\right)^2,\\
\end{aligned}
\end{equation}
where $R_s\equiv 2GM$ is the horizon radius of the Schwarzschild black hole. From Eq.~(\ref{heff}) and Eq.~(\ref{fr}), it is clear that when $\gamma \rightarrow 0$, $H_{\mathrm{eff}}^{2}\rightarrow \frac{8\pi G}{3}\rho-\frac{k}{a^2}$ returns to the classical case, and $f\left( r \right) \rightarrow 1-\frac{R_s}{r}$ returns to the Schwarzschild metric. When $k=0$, Eq.~(\ref{fr}) is consistent with the metric in \cite{lewandowski2023quantum}.

To study the thermodynamic behavior of this quantum-corrected black hole, we extend it to AdS spacetime, obtaining the line element of the quantum-corrected-AdS black hole as
\begin{equation}
	f(r)=1-\frac{R_s}{r}+\frac{4\sqrt{3}\pi\gamma^3G\hbar}{r^2}\left(\frac{R_s}{r}-k\tilde{r}_0^2\right)^2-\frac{\varLambda r^2}{3}.
\end{equation}
Using natural units $(c=G=\hbar=1)$, this becomes
\begin{equation}
	f\left( r \right) =1-\frac{2M}{r}+\frac{4\sqrt{3}\pi \gamma ^3}{r^2}\left( \frac{2M}{r}-k\tilde{r}_0^2 \right) ^2-\frac{\varLambda r^2}{3}.
\end{equation}

\section{Thermodynamics of quantum-corrected-AdS black hole in extended phase space}\label{sec3}

The radius $r_{+}$ of the outer black hole event horizon is the largest root of $f(r_{+})=0$, which can be obtained by solving
\begin{equation}
	\frac{\varLambda}{3}{r^6_+}-{r^4_+}+2{Mr^3_+}+4\sqrt{3}\pi \gamma ^3k^2\tilde{r}_0^4{r^2_+}+16\sqrt{3}\pi \gamma ^3k\tilde{r}_0^2Mr_+-16\sqrt{3}\pi \gamma ^3M^2=0.
\end{equation}
This is a quadratic equation in $M$, leading to
\begin{equation}
	M_{\pm}=\frac{\sqrt{3}{r^3_+}+24\pi \gamma ^3k\tilde{r}_0^2r_+\pm \sqrt{\left( 16\sqrt{3}\pi \gamma ^3\varLambda +3 \right) {r^6_+}+48\sqrt{3}\pi \gamma ^3\left( k\tilde{r}_0^2-1 \right) {r^4_+}}}{48\pi \gamma ^3}.
\end{equation}
$M_+$ is discarded because $\underset{\gamma \rightarrow 0}{\lim}M_+\rightarrow \infty$, and $\underset{\gamma \rightarrow 0}{\lim}M_-\rightarrow M_{Sch-AdS}$ (the Schwarzschild-AdS black hole case).
Using the specific volume $v=2r_+$, the pressure $P=-\frac{\Lambda}{8\pi}$, and $M_-$, the Hawking temperature of the black hole is defined as
\begin{equation}
	\begin{aligned}
	T&=\frac{f^{\prime}(r_+)}{4\pi}=\frac{k\tilde{r}_0^2\sqrt{\left( 9-384\sqrt{3}\pi ^2\gamma ^3P \right) v^2-576\sqrt{3}\pi \gamma ^3\left( 1-k\tilde{r}_0^2 \right)}}{6\pi v^2}\\
	&+\frac{2\pi Pv^2+2\left( 1-k\tilde{r}_0^2 \right)}{\pi v}
	+\frac{-\sqrt{3}v+\sqrt{\left( 3-128\sqrt{3}\pi ^2\gamma ^3P \right) v^2-192\sqrt{3}\pi \gamma ^3\left( 1-k\tilde{r}_0^2 \right)}}{64\pi ^2\gamma ^3}.
	\end{aligned}
\end{equation}
From this equation, the pressure of the black hole can be solved as (similarly discarding the divergent branch when $\gamma\rightarrow 0$)
\begin{equation}\label{pvt}
	\begin{aligned}
		P(v,T)&=-[\sqrt{3}(\frac{9}{\gamma ^6}v^8-\frac{768\sqrt{3}\pi ^2}{\gamma ^3}Tv^7-\frac{384\sqrt{3}\pi \left( 2-3k\tilde{r}_0^2 \right)}{\gamma ^3}v^6\\
		&-49152\pi ^3k\tilde{r}_0^2Tv^5+6144\pi ^2k\tilde{r}_0^2\left( 11k\tilde{r}_0^2-8 \right) v^4-262144\sqrt{3}\pi ^4\gamma ^3k^2\tilde{r}_0^4Tv^3\\
		&-131072\sqrt{3}\pi ^3\gamma ^3k^2\tilde{r}_0^4\left( 2-3k\tilde{r}_0^2 \right) v^2+1048576\pi ^4\gamma ^6k^4\tilde{r}_0^8)^{\frac{1}{2}}\\
		&-1024\sqrt{3}\pi ^2\gamma ^3k^2\tilde{r}_0^4]\frac{1}{768\pi ^2v^4}+\frac{3k\tilde{r}_0^2-4}{4\pi v^2}+\frac{T}{2v}+\frac{\sqrt{3}}{256\pi ^2\gamma ^3}.
    \end{aligned}
\end{equation}

\begin{figure}[h]
	\centering
	\includegraphics[width=1\textwidth]{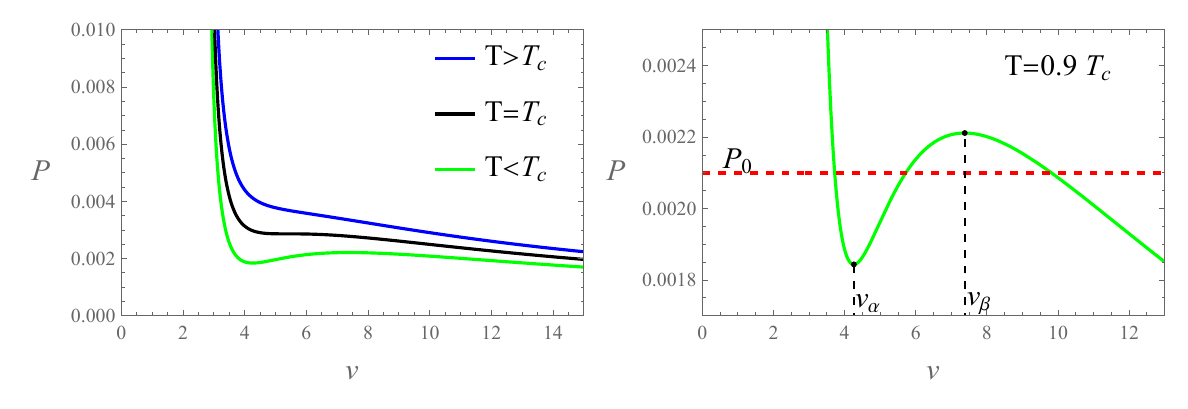}
	\caption{The $P-v$ isothermal curves for quantum-corrected-AdS black hole with fixed parameters $k = 0$, $\tilde{r}_0 = 0.5$, and $\gamma = 0.2375$ at different temperatures $T$. The blue, black, and green lines correspond to $T = 1.1T_c$, $T = T_c$, and $T = 0.9T_c$, respectively. In the right panel with $T = 0.9T_c$, the red dashed line represents the coexistence line of large and small black holes at the phase transition pressure $P_0$. $v_{\alpha}$ and $v_{\beta}$ are the positions of the two extrema on the curve.}
	\label{fdwes}
\end{figure}

Fig.~\ref{fdwes} shows that a large-small black hole phase transition exists, which is qualitatively similar to the gas-liquid phase transition in the Van der Waals system. Specifically, when the temperature $T$ is less than the critical temperature $T_c$, the $P-v$ curve divides into three branches. The left ($v<v_{\alpha}$) branch corresponds to small black holes, the middle ($v_{\alpha}<v<v_{\beta}$) branch corresponds to intermediate black holes, and the right ($v>v_{\beta}$) branch corresponds to large black holes. When studying chaos, we assume the system is within the interval $(v_{\alpha},v_{\beta})$, known as the spinodal region. The two points $v_{\alpha}$ and $v_{\beta}$ are determined by $\left. \partial _vP\left( v,T_0 \right) \right|_{v_{\alpha}}=\left. \partial _vP\left( v,T_0 \right) \right|_{v_{\beta}}=0$. According to the heat capacity, branches with negative slopes represent thermodynamically stable black holes, while branches with positive slopes represent thermodynamically unstable black holes. Therefore, small and large black holes are stable, while intermediate black holes are unstable. Using the Maxwell equal area law, the phase transition points and the phase transition pressure $P_0$ on the $P-v$ curve can be determined.

The critical point can be determined by the following conditions
\begin{equation}
	(\partial_{v}P)_{T,k,\tilde{r}_0,\gamma}=(\partial_{v,v}P)_{T,k,\tilde{r}_0,\gamma}=0.
\end{equation}
For $k=0$, the critical point has an analytical solution given by
\begin{equation}\label{ptv}
	\begin{gathered}
		T_{c}=\frac{\sqrt{3\sqrt{3}}}{40\sqrt{5}\pi^{\frac{3}{2}}\gamma^{\frac{3}{2}}}, \\
		P_{c}=\frac{7\sqrt{3}}{32000\pi^{2}\gamma^{3}}, \\
		v_{c}=\frac{80\sqrt{5}\pi^{\frac{1}{2}}\gamma^{\frac{3}{2}}}{3\sqrt{3\sqrt{3}}}. 
    \end{gathered}
\end{equation}
It is obvious that the three critical values $T_c$, $P_c$, and $V_c$ are only related to the parameter $\gamma$ (but are also related to $\tilde{r}_0$ for $k=\pm 1$). Using Eq.~(\ref{ptv}), a critical ratio $\frac{P_cV_c}{T_c}=\frac{7}{18}=0.38\dot{8}$ can be defined. The results indicate that this ratio is independent of the parameters $\gamma$ and $\tilde{r}_0$. For $k=\pm 1$, although there are no analytical solutions, numerical analysis shows that this ratio is also independent of $\gamma$ but dependent on $\tilde{r}_0$. Interestingly, for the classical Van der Waals system and most AdS black holes, this ratio is $\frac{3}{8}=0.375$. 

\begin{figure}[h]
	\centering
	\subfigure{\includegraphics[width=1\linewidth]{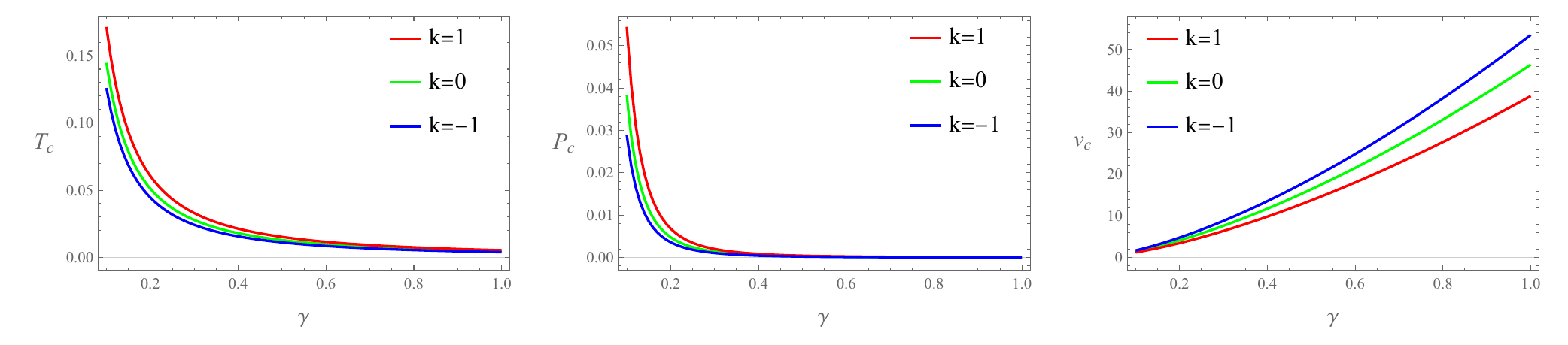}}
	\hfill
	\subfigure{\includegraphics[width=1\linewidth]{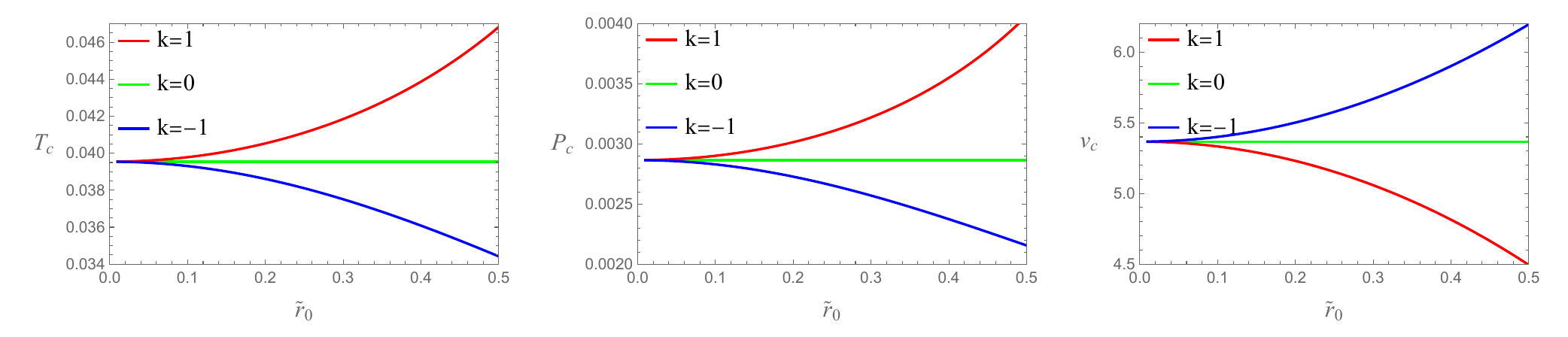}}
	\caption{The variations of $T_c$, $P_c$, and $V_c$ with $\gamma$ (for fixed $\tilde{r}_0=0.5$) and with $\tilde{r}_0$ (for fixed $\gamma=0.2375$) in the three types of universes.}
	\label{TPVc}
\end{figure}

Fig.~\ref{TPVc} shows the variations of $T_c$, $P_c$, and $V_c$ with respect to the parameters $\gamma$ and $\tilde{r}_0$. 
As shown in the figure, with the increase of the parameter $\gamma$, both $T_c$ and $P_c$ monotonically decrease, while $V_c$ monotonically increases. A larger cosmic curvature $k$ leads to higher values of $T_c$ and $P_c$. However, as $\gamma$ increases, the red, green, and blue curves gradually converge and become difficult to distinguish, whereas the behavior of $V_c$ is the opposite. In different universes, $T_c$, $P_c$, and $V_c$ coincide at $\tilde{r}_0=0$ because $\tilde{r}_0=0$ causes $k$ to disappear from the metric. As the parameter $\tilde{r}_0$ increases, $T_c$, $P_c$, and $V_c$ begin to diverge. Specifically, $T_c$, $P_c$, and $V_c$ remain unchanged in the flat universe since $k=0$ causes $\tilde{r}_0$ to disappear from the metric. $T_c$ and $P_c$ increase monotonically in the closed universe and decrease monotonically in the open universe, while $V_c$ shows the opposite behavior.

\begin{figure}[h]
	\centering
	\includegraphics[width=0.6\textwidth]{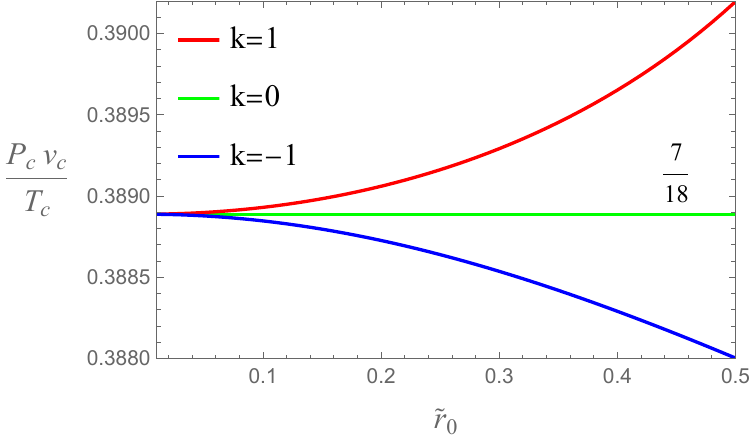}
	\caption{The variations of $T_c$, $P_c$, and $V_c$ with $\tilde{r}_0$ in the three types of universes (independent of the value of the quantum correction parameter$\gamma$).}
	\label{cz}
\end{figure}

Fig.~\ref{cz} shows the variation of $\frac{P_cV_c}{T_c}$ with the parameter $\tilde{r}_0$. Although Fig.~\ref{TPVc} indicates that $T_c$, $P_c$, and $V_c$ are related to $\gamma$ in all three types of universes, $\gamma$ subtly disappears from $\frac{P_cV_c}{T_c}$, as shown in Eq.~(\ref{ptv}). Similarly, at $\tilde{r}_0=0$, $\frac{P_cV_c}{T_c}=\frac{7}{18}$ in all three types of universes. As $\tilde{r}_0$ increases, $\frac{P_cV_c}{T_c}$ remains constant in a flat universe, increases monotonically in a closed universe, and decreases monotonically in an open universe. For the Van der Waals system, where $\frac{P_cV_c}{T_c}=\frac{3}{8}$, Fig.~\ref{cz} shows that only an open universe can attain this value, while $\frac{P_cV_c}{T_c}$ is always greater in the other two universes.
Different black holes can originate from collapsing dust spheres with varying radii $\tilde{r}_0$. Therefore, the trend of $\frac{P_cV_c}{T_c}$ varying with $\tilde{r}_0$ might be used to validate the quantum correction model studied in this paper and to determine the curvature of our universe.

We briefly discussed the thermodynamics of this quantum-corrected-AdS black hole. More detailed discussions are available in \cite{wang2024thermodynamics}. Next, we will use the Melnikov method to study the chaotic behavior of this black hole under periodic thermal perturbations.

\section{Chaos in the quantum-corrected-AdS black hole flow under thermal perturbations}\label{sec4}

We adopt the same approach as in \cite{slemrod1985temporal} to study chaotic phenomena in "black hole fluids." We assume the "black hole fluid" to be a one-dimensional fluid column of fixed volume flowing along the $x$-direction. We clarify that this assumption does not necessarily represent the physical reality of black hole fluids. Given that black holes exhibit thermodynamic properties similar to those of the Van der Waals system, we hope this assumption is valid and can reveal the thermodynamic chaotic behavior of black holes.

\subsection{Temporal chaos in spinodal region}\label{subsec1}

We first study the effects of time-periodic perturbations on the black hole fluid system. The position of a reference fluid particle in the black hole fluid is represented by the Eulerian coordinate $x_0$. The mass $M$ of the fluid column with a unit cross-section between the Eulerian coordinates $x_0$ and $x$ of general fluid particles is
\begin{equation}\label{M}
	M(x,t)=\int_{x_0}^x\rho(\xi,t)d\xi,
\end{equation}
where $\rho(\xi,t)$ is the fluid density at spatial position $x$ and time $t$. Through Eq.~(\ref{M}), the spatial position $x$ can be obtained as $x=x(M,t)$. Furthermore, we define $v\left( M,t \right) :=x_M\left( M,t \right) \equiv \partial _Mx\left( M,t \right) =\rho \left( x,t \right) ^{-1}$ and $u\left( M,t \right) :=x_t\left( M,t \right) \equiv \partial _tx\left( M,t \right)$, which represent the specific volume and velocity, respectively. In this case the equations of balance of mass and momentum are, respectively, 
\begin{equation}\label{vt}
	\frac{\partial v}{\partial t}=\frac{\partial u}{\partial M},
\end{equation}
\begin{equation}\label{ut}
	\frac{\partial u}{\partial t}=\frac{\partial\tau}{\partial M},
\end{equation}
where $\tau$ is the Piola stress. By assuming the fluid is thermoelastic, slightly viscous, and isotropic, the Piola stress takes a form as
\begin{equation}\label{tau}
	\tau=-P(v,T)+\mu u_M-Av_{MM}.
\end{equation}
where $A$ is a positive constant and $\mu$ is a small positive constant viscosity. Substituting Eq.~(\ref{tau}) into Eq.~(\ref{ut}) yields
\begin{equation}\label{xtt}
	x_{tt}=-P(v,T)_{M}+\mu u_{MM}-Av_{MMM}.
\end{equation}
To simplify subsequent calculations, we transform $M$, $x$, and $t$ as follows: $\tilde{M}=sM$, $\tilde{x}=sx$, and $\tilde{t}=st$ with $s=\frac{2 \pi}{M}$, which makes the ranges of $\tilde{M}$, $\tilde{x}$, and $\tilde{t}$ all lie within $[0, 2\pi]$. Eq.~(\ref{xtt}) can then be rewritten as (with overbars deleted)
\begin{equation}\label{newxtt}
	x_{tt}=-P(v,T)_{M}+\epsilon\mu_{0}su_{MM}-As^{2}v_{MMM},
\end{equation}
where $\mu\equiv \epsilon\mu_{0}$ with $0<\epsilon\ll 1$ and $\mu_{0}>0$. Next, we use the tools of Hamiltonian mechanics to systematically analyze the dynamical behavior and stability of the system. Eq.~(\ref{newxtt}) can be reformulated into the form of a dissipative Hamiltonian system given by
\begin{equation}\label{xu}
	\begin{aligned}
	&x_t=\partial _uH\left( x,u;T \right),\\
	&u_t=-\partial _xH\left( x,u;T \right) +\epsilon \mu _0su_{MM},
	\end{aligned}
\end{equation}
where
\begin{equation}\label{H}
	H(x,u;T)=\frac{1}{\pi}\int_{0}^{2\pi}\left[\frac{u^{2}}{2}-\int_{v_{0}}^{v}P(v,T)dv+\frac{As^{2}}{2}v_{M}^{2}\right]dM.
\end{equation}

Now we assume that the fluid system is in the unstable spinodal region where small and large black holes coexist, with the specific volume $v=v_0$ (the
inflection point, satisfying $\partial _{v,v}P\left( v,T_0 \right) |_{v_0}=0$) and temperature $T=T_{0}<T_c$. Taking this as the initial state of the system's evolution, we introduce a time-periodic perturbation in the form
\begin{equation}\label{twr}
	T=T_0+\epsilon\delta\cos\omega t\cos M,
\end{equation}
where $\epsilon$ is consistent with the $\epsilon$ in Eq.~(\ref{newxtt}), $\delta$ represents the amplitude of the perturbation, which is related to the viscosity of the fluid system, and $\omega$ is the fluctuation angular frequency. To perform a perturbation analysis on the system, we expand $P(v,T)$ in a Taylor series around $(v_0,T_0$) as (truncate at cubic terms)
\begin{equation}\label{ptl}
	\begin{aligned}P(v,T)&=P(v_{0},T_{0})+P_{v}(v_{0},T_{0})(v-v_{0})+P_{T}(v_{0},T_{0})(T-T_{0})\\&+P_{vT}(v_{0},T_{0})(v-v_{0})(T-T_{0})+P_{TT}(v_{0},T_{0})(T-T_{0})^2\\
		&+\frac{P_{vvv}(v_{0},T_{0})}{3!}(v-v_{0})^{3}+\frac{P_{TTT}(v_{0},T_{0})}{3!}(T-T_{0})^{3}\\
		&+\frac{P_{vvT}(v_{0},T_{0})}{2}(v-v_{0})^{2}(T-T_{0})+\frac{P_{vTT}(v_{0},T_{0})}{2}(v-v_{0})(T-T_{0})^2,
	\end{aligned}
\end{equation}
where $P_{vv}=0$ because $v=v_0$ is the inflection point of the $P-v$ curve. Unlike in other papers, since $P(v,T)$ in Eq.~(\ref{pvt}) is not a linear function of $T$, $P_{TT}(v_{0},T_{0})$, $P_{TTT}(v_{0},T_{0})$, and $P_{vTT}(v_{0},T_{0})$ are all non-zero. However, we will soon see that these additional non-zero terms only appear in the higher-order terms of $\epsilon$. After neglecting these higher-order terms, their contribution is zero.

At the same time, $v(M,t)$ and $u(M,t)$ can be expanded, respectively, as Fourier cosine and sine series in terms of $M$ as
\begin{equation}\label{vutl}
	\begin{aligned}
	&v(M,t)=x_{M}(M,t)=v_{0}+x_{1}(t)\cos M+x_{2}(t)\cos2M+x_{3}(t)\cos3M+\cdots,\\
	&u(M,t)=x_{t}(M,t)=u_{1}(t)\sin M+u_{2}(t)\sin2M+u_{3}(t)\sin3M+\cdots,
    \end{aligned}
\end{equation}

where $x_i(t)$ represents the displacement modes of the system, and $u_i(t)$ represents the velocity modes. These describe the hydrodynamic behavior of the system deviating from the initial equilibrium state $v=v_0$. Previous studies have shown that primary modes often capture the core dynamic characteristics of the system, providing sufficient physical insights. For example, Slemrod and Marsden, in their study of van der Waals fluid dynamics, successfully revealed the key chaotic behavior of the system by simplifying to primary modes \cite{slemrod1985dynamics}. Similarly, Holmes, in analyzing the chaotic behavior of nonlinear systems, found that higher-order modes had a minimal impact on the main dynamic characteristics \cite{holmes1990chaotic}. Therefore, we choose to focus on the primary modes $x_1(t)$ and $u_1(t)$, which not only simplifies the analysis process but also allows us to effectively explore the core dynamic characteristics of the system.

Considering these, substituting Eq~.(\ref{twr}), (\ref{ptl}), and (\ref{vutl}) into Eq~.(\ref{H}) leads to (omitting the subscript 1)
\begin{equation}\label{HH}
	\begin{aligned}
	H\left( x,u \right) &=\frac{1}{2}u^2+\frac{1}{2}\left( As^2-P_v \right) x^2-\frac{P_{vvv}}{32}x^4-\epsilon \delta \cos \omega t\,\left( P_Tx+\frac{P_{vvT}}{8}x^3 \right)\\
	& -\epsilon ^2\delta ^2\cos ^2wt\frac{3P_{vTT}}{16}x^2-\epsilon ^3\delta ^3\cos ^3wt\frac{P_{TTT}}{8}x,
	\end{aligned}
\end{equation}
and then substituting Eq~.(\ref{HH}) into Eq~.(\ref{xu}) finally gives
\begin{equation}\label{xuu}
	\begin{aligned}
		&x_t\equiv\dot{x}=u,\\
		&\begin{aligned}
		u_t\equiv\dot{u}&=(P_{v}-As^{2})x+\frac{P_{vvv}}{8}x^{3}+\epsilon\delta\cos\omega t\left(P_{T}+\frac{3P_{vvT}}{8}x^{2}\right)-\epsilon\mu_{0}su\\
		&+\epsilon^2\delta^2\cos^2\omega t\frac{3P_{vTT}}{8}x+\epsilon^3\delta^3\cos^3\omega t\frac{P_{TTT}}{8}.
		\end{aligned}
	\end{aligned}
\end{equation}
Since $\epsilon \ll 1$, we can ignore the higher-order terms of $\epsilon$, making $\dot{u}$ consistent with the $\dot{u}$ in \cite{slemrod1985temporal}. Setting $z\equiv[x,u]^{T}$, Eq.~(\ref{xu}) can be rewritten as
\begin{equation}\label{zd}
	\dot z=f(z)+\epsilon g(z,t),
\end{equation}
with
\begin{equation}\label{fz}
	f(z)=\begin{bmatrix}u\\a^2x+\frac{P_{vvv}}{8}x^3\end{bmatrix},
\end{equation}
and
\begin{equation}\label{gz}
	g(z)=\begin{bmatrix}0\\\left(P_T+\frac{3P_{vvT}}{8}x^2\right)\delta\cos\omega t-\mu_0su\end{bmatrix},
\end{equation}
where $a^2\equiv(P_{v}-As^{2})$. For the case without time-periodic perturbation ($\epsilon=0$), the analytical solution to Eq.~(\ref{zd}) is \cite{holmes1979nonlinear,slemrod1985temporal}
\begin{equation}\label{zl}
	z_0(t)=\begin{bmatrix}x_0(t)\\u_0(t)\end{bmatrix}=\begin{bmatrix}\frac{\pm4a}{(-P_{vvv})^{1/2}}\mathrm{sech}(at)\\\frac{\mp4a^2}{(-P_{vvv})^{1/2}}\mathrm{sech}(at)\tanh(at)\end{bmatrix}.
\end{equation}

\begin{figure}[h]
	\centering
	\includegraphics[width=0.5\textwidth]{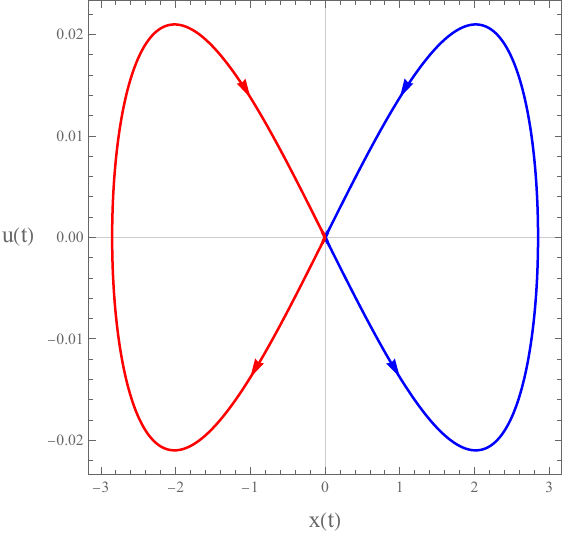}
	\caption{The homoclinic orbit described by Eq.~(\ref{zl}), with the arrows indicating the increasing direction of time $t$. We fixed the parameters as follows: $k=0$, $\gamma=0.2375$, $T=0.9T_c\approx0.03559$, $A=0.2$, and $s=0.001$.}
	\label{sjhdd}
\end{figure}

As shown in Fig.~\ref{sjhdd}, in the phase space $(x,u)$, Solution (\ref{zl}) describes a homoclinic orbit connecting a saddle point (the origin) to itself. The two branches of the solution correspond to the left and right parts of the orbit. When $t\rightarrow \pm \infty$, $z_0$ approaches the saddle point.

For the case with time-periodic perturbation ($\epsilon \ne 0$), we use the Melnikov method to determine the chaotic behavior of the black hole fluid system. Consider the perpendicular  distance from the stable manifold to the unstable manifold on the Poincaré surface at $t=t_0$ to be

\begin{equation}\label{dt}
	d(t_0)=\frac{\varepsilon M(t_0)}{|f(\mathbf{z}_0(0))|},
\end{equation}
where $M(t_0)$ is the Melnikov function, which takes the form
\begin{equation}\label{Mt}
	M\left(t_0\right)=\int_{-\infty}^{+\infty}f^{\mathrm{T}}\left[z_0\left(t-t_0\right)\right]\mathbf{J}g\left[z_0\left(t-t_0\right)\right]dt,
\end{equation}
with
\begin{equation}
	\mathbf{J}=\begin{bmatrix}0&1\\-1&0\end{bmatrix}.
\end{equation}
If the Melnikov function $M(t_0)$ given by Eq.~(\ref{Mt}) is zero at $t=t_0$, then the perpendicular distance $d(t_0)$ defined by Eq.~(\ref{dt}) will also be zero. This means that the stable and unstable manifolds intersect transversely at $t=t_0$, resulting in a Smale horseshoe in the Poincaré map, indicating the onset of chaos. Therefore, by solving the system's Melnikov function $M(t_0)$ and determining whether it has zeros, we can ascertain if the system exhibits chaotic behavior.

\begin{figure}[h]
	\centering
	\includegraphics[width=1\textwidth]{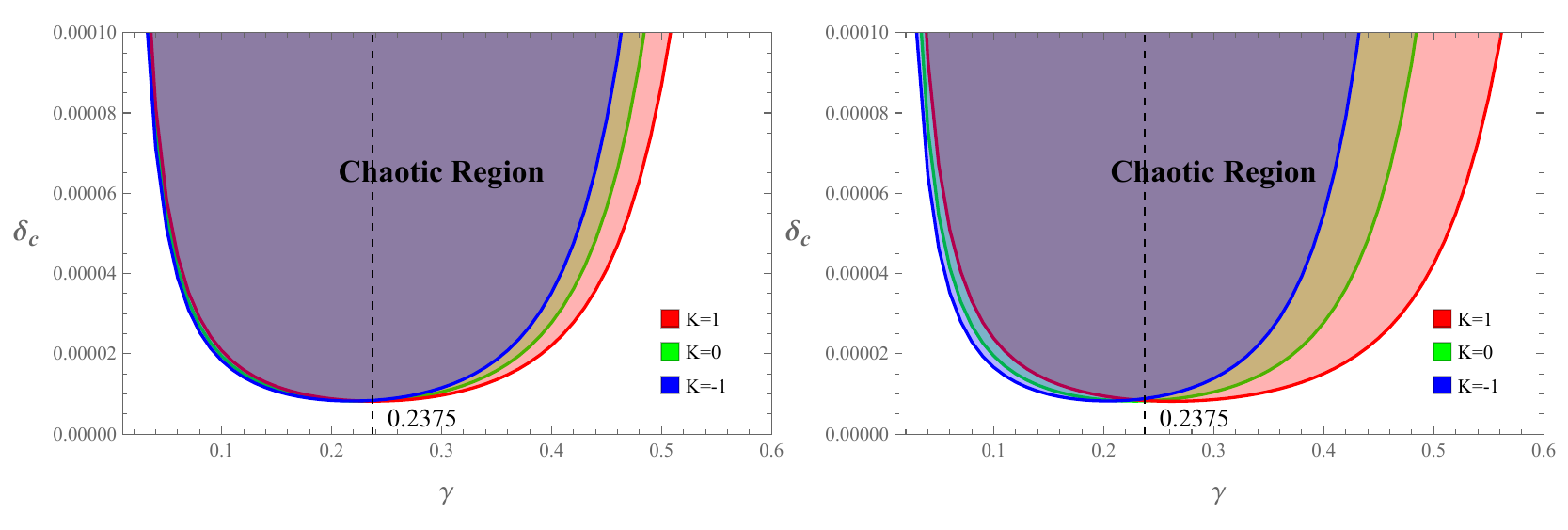}
	\caption{The left and right panels respectively show the variation of the critical amplitude $\delta_{c}$ with the quantum correction parameter $\gamma$ for $\tilde{r}_0=0.3$ and $\tilde{r}_0=0.5$. The shaded regions in the figures represent the areas where chaos occurs. We fixed $T=0.9T_c(k)$, $A=0.2$, $s=0.001$, $\omega=0.01$, $\epsilon=0.001$, and $\mu_{0}=0.1$.}
	\label{chaotic}
\end{figure}

Now, by substituting Eq.~(\ref{fz}), (\ref{gz}), and (\ref{zl}) into Eq.~(\ref{Mt}), we obtain
\begin{equation}
	\begin{aligned}
		M(t_{0})& =\int_{-\infty}^{\infty}dt\frac{4a^{2}}{(-P_{vvv})^{1/2}}\frac{\delta\cos\omega t\sinh a(t-t_{0})}{\cosh^{2}a(t-t_{0})}\left[\frac{6a^2P_{vvT}}{P_{vvv}}\mathrm{sech}^2a(t-t_0)-P_T\right] \\
		&+\int_{-\infty}^{\infty}dt\frac{16\mu_{0}sa^{4}}{P_{vvv}}\mathrm{sech}^{2}a(t-t_{0})\tanh^{2}a(t-t_{0}).
    \end{aligned}
\end{equation}
Using the residue theorem to compute the integral of the above expression, $M(t_0)$ can be further expressed as
\begin{equation}
	M(t_0)=\delta\omega K\sin\omega t_0+\mu_0sL,
\end{equation}
with
\begin{equation}
	\begin{aligned}
		K=\frac{8\pi}{(-P_{vvv})^{1/2}}\left[P_{T}-\frac{P_{vvT}}{P_{vvv}}(\omega^{2}+a^{2})\right]\frac{\exp(\frac{\pi\omega}{2a})}{1+\exp(\frac{\pi\omega}{a})},\quad L=\frac{32a^3}{3P_{vvv}}.
	\end{aligned}
\end{equation}
It is then easy to see that $M(t_0)$ has a simple zero if $\left|\frac{s\mu_0L}{\delta\omega K}\right|\leq1$, which implies that $\delta$ has a critical value satisfying
\begin{equation}
	\delta_{c}=\left|\frac{s\mu_{0}L}{\omega K}\right|.
\end{equation}
This indicates that chaos occurs when $\delta>\delta_{c}$.
Fig.~\ref{chaotic} shows the variation of the critical value $\delta_{c}$ with respect to the parameter $\gamma$. It can be observed that when $\gamma$ is very small, which can be regarded as approaching the case of a Schwarzschild black hole, chaos is less likely to occur. As $\gamma$ increases to around 0.2375, chaos becomes more likely. Further increasing $\gamma$ makes chaos less likely to occur again. This phenomenon holds true in universes with all three values of $k$. It should be noted that the increased likelihood of chaos at $\gamma=0.2375$ is coincidental. In fact, we set several different perturbation parameters and found that the concave part of the $\delta_{c}-\gamma$ curve does not necessarily concentrate around $\gamma=0.2375$. However, it is certain that chaos is more likely to occur at $\gamma=0.2375$ than at $\gamma=0$. Therefore, based on the $\gamma=0.2375$ identified in Reference [1], the quantum-corrected model studied here makes the black hole fluid system more prone to chaos. On the other hand, comparing the left ( $\tilde{r}_0=0.3$ ) and right ( $\tilde{r}_0=0.5$ ) panels of Fig.~\ref{chaotic}, it is evident that as $\tilde{r}_0$ increases from $0$, the $\delta_{c}-\gamma$ curves for the three different universes start to diverge (at $\tilde{r}_0$, $k$ disappears from the metric, causing the $\delta_{c}-\gamma$ curves of the three universes to coincide). Specifically, for the flat universe ($k=0$), the $\delta_{c}-\gamma$ curve remains unchanged, which is reasonable because $k=0$ also causes $\tilde{r}_0$ to disappear from the metric. For the closed universe ($k=1$), $\delta_{c}$ increases at smaller $\gamma$ (making chaos less likely to occur) and decreases at larger $\gamma$ (making chaos more likely to occur). The open universe ($k=-1$) exhibits the opposite behavior to the closed universe. These changes imply that the $\delta_{c}-\gamma$ curves for any two different universes must intersect at some point, which is an interesting phenomenon.

\begin{figure}[h]
	\centering
	\includegraphics[width=0.6\textwidth]{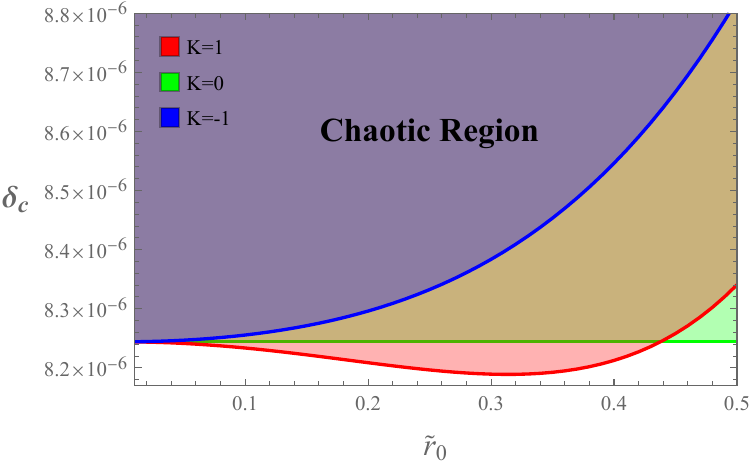}
	\caption{The variation of $\delta_{c}$ with $\tilde{r}_0$ when $\gamma=0.2375$. The shaded regions in the figures represent the areas where chaos occurs. We fixed $T=0.9T_c(k)$, $A=0.2$, $s=0.001$, $\omega=0.01$, $\epsilon=0.001$, and $\mu_{0}=0.1$.}
	\label{dr}
\end{figure}

This phenomenon results in different changes in $\delta_{c}$ with $\tilde{r}_0$ for different values of $\gamma$. As an example, we plot the $\delta_{c}-\tilde{r}_0$ curves for $\gamma=0.2375$ in Fig.~\ref{dr}. It is easy to see that at $\tilde{r}_0=0$, the $\delta_{c}$ values for the three types of universes coincide because $k$ disappears from the metric at this point. As $\tilde{r}_0$ increases, $\delta_{c}$ remains unchanged in the flat universe, decreases first and then increases in the closed universe, and increases continuously in the open universe. This actually depends on the position of the intersection points of the $\delta_{c}-\gamma$ curves for the closed and open universes with the $\delta_{c}-\gamma$ curve for the flat universe. As $\tilde{r}_0$ increases, the $\gamma$-coordinate of the intersection point for the closed universe moves from left to right and passes through $\gamma=0.2375$, while the intersection point for the open universe moves from right to left and always remains to the left of $\gamma=0.2375$. Overall, for the real scenario (if $\gamma=0.2375$ is entirely accurate), the variations of $\delta_{c}$ with $\tilde{r}_0$ in different universes are completely distinct. This helps in distinguishing which type of universe ours belongs to.

\begin{figure}[h]
	\centering
	\includegraphics[width=1\textwidth]{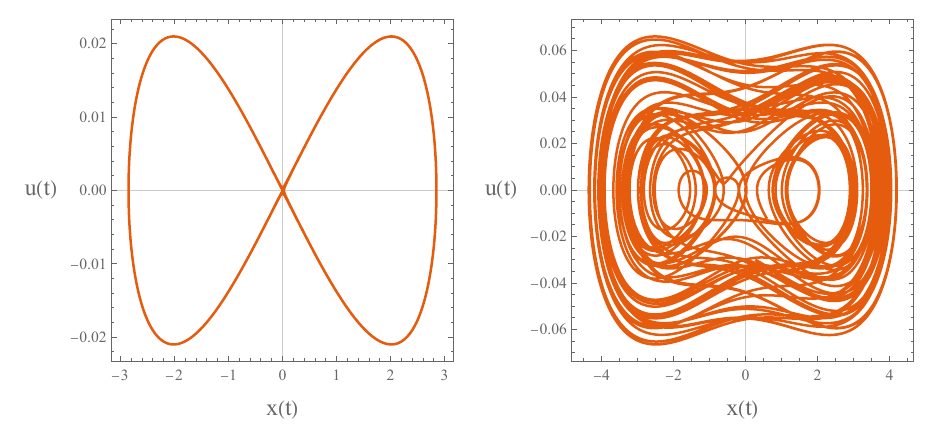}
	\caption{On the left (for $\delta=1\times 10^{-7}<\delta_c\approx8.2441\times 10^{-6}$) and the right (for $\delta=1>\delta_c\approx8.2441\times 10^{-6}$) are the evolution phase diagrams of the black hole fluid system under time-periodic perturbations. We fixed the parameters as follows: $k=0$, $\gamma=0.2375$, $T\approx0.03559=0.9T_c$, $A=0.2$, $s=0.001$, $\omega=0.01$, $\epsilon=0.001$, and $\mu_{0}=0.1$.}
	\label{sjhd}
\end{figure}

Fig.~\ref{sjhd} shows the evolution phase diagram of the black hole fluid system in the $x(t)-u(t)$ phase space for $k=0$ and $\gamma=0.2375$, as determined by Eq.~(\ref{xu}). The results indicate that when $\delta<\delta_{c}$, the system's evolution trajectory is very regular, and the fluid eventually approaches an equilibrium state with $v=v_0$. When $\delta>\delta_{c}$, the system's evolution trajectory becomes highly irregular, exhibiting chaotic behavior.

\subsection{Spatial chaos in the equilibrium state}\label{subsec2}

\begin{figure}[h]
	\centering
	\includegraphics[width=1\textwidth]{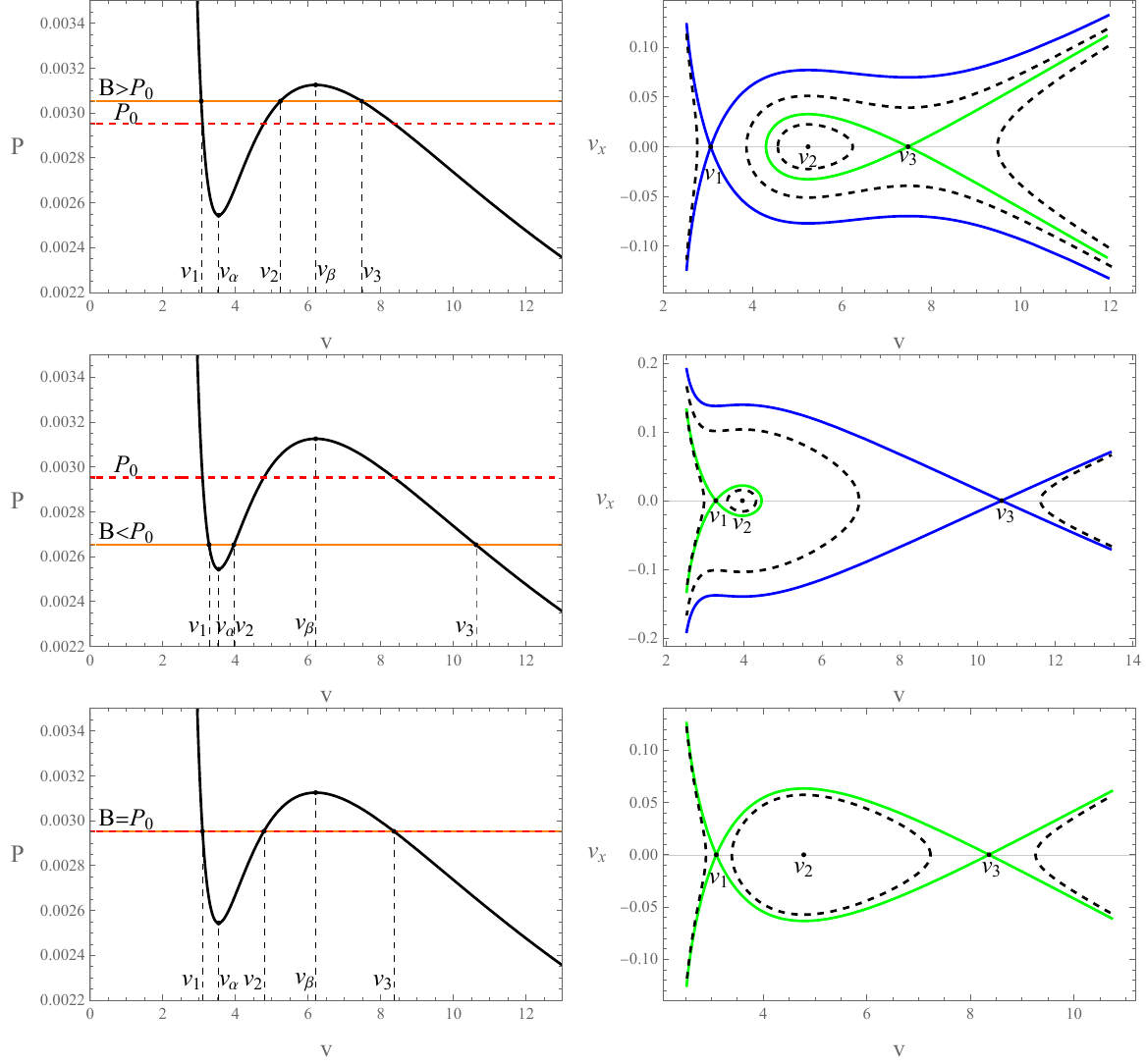}
	\caption{The figure shows the $P-v$ isotherms (left side) and the corresponding unperturbed phase diagrams in the $v-v_x$ phase space (right side) for a closed universe ($k=1$). From top to bottom, they correspond to $P_0<B<P(v_{\beta},T_0)$, $P(v_{\alpha},T_0)<B<P_0$, and $B=P_0$, respectively. We set $\gamma=0.2375$, $\tilde{r}_0=0.5$, $T=0.9 T_c\approx0.04215$, and $A=0.2$.}
	\label{PV1}
\end{figure}

\begin{figure}[h]
	\centering
	\includegraphics[width=1\textwidth]{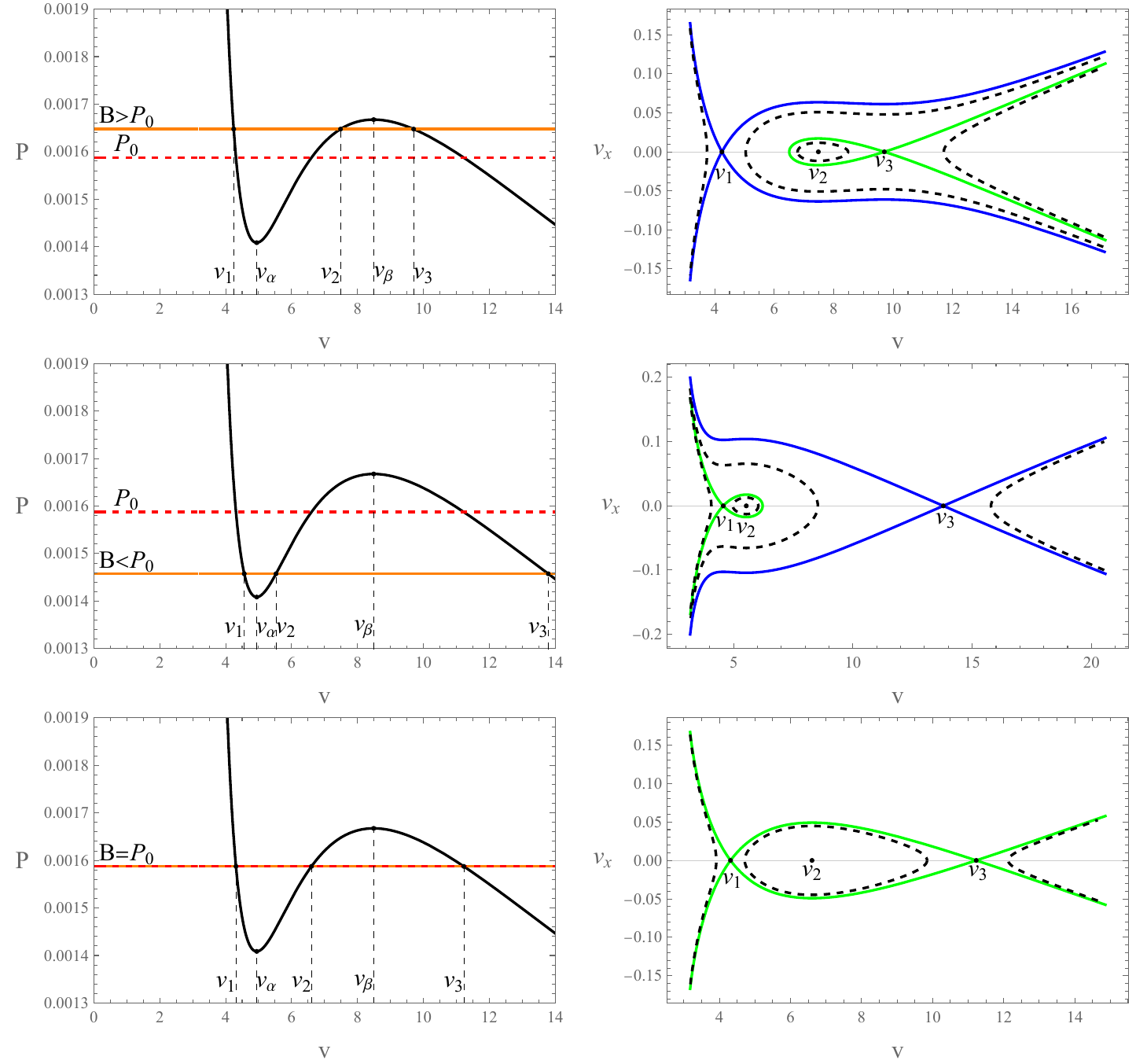}
	\caption{The figure shows the $P-v$ isotherms (left side) and the corresponding unperturbed phase diagrams in the $v-v_x$ phase space (right side) for a open universe ($k=-1$). From top to bottom, they correspond to $P_0<B<P(v_{\beta},T_0)$, $P(v_{\alpha},T_0)<B<P_0$, and $B=P_0$, respectively. We set $\gamma=0.2375$, $\tilde{r}_0=0.5$, $T=0.9 T_c\approx0.03098$, and $A=0.2$.}
	\label{PV2}
\end{figure}

In this section, we investigate the chaotic phenomena in the black hole fluid system induced by spatially periodic perturbations. Assuming the black hole fluid system is in a thermal equilibrium state at a temperature $T<T_c$, its state is described by Eq.~(\ref{pvt}). According to the van der Waals–Korteweg theory, the external force $\tau$ in Eq.~(\ref{tau}) should be rewritten as
\begin{equation}\label{tauu}
	\tau=-P(v,T_{0})-Av^{''},
\end{equation}
where $A$ is a positive constant, and the prime symbol $\prime$ denotes differentiation with respect to the Eulerian coordinate $x$. In thermal equilibrium, the black hole fluid system has no body forces, satisfying $\tau^{\prime}=0$. Thus, $\tau=B$ is constant, where this constant represents the ambient pressure at the end of the fluid column. Substituting $\tau=B$ into Eq.~(\ref{tauu}) gives
\begin{equation}\label{BB}
	Av''+P(v,T_0)=B,\quad-\infty<x<\infty.
\end{equation}
For a thermodynamic system at $T_0<T_c$, using the Maxwell equal area rule, we can obtain the pressure $P_0$ associated with the phase transition between small and large black holes. At the same time, the nonlinear system in Eq.~(\ref{BB}) has three fixed points, which are the specific volumes $v_1$, $v_2$, and $v_3$ at the three intersection points of $P=B$ and the $P-v$ curve. When there are no perturbations, the evolution state of the black hole fluid system can be classified into three scenarios based on the relationship between the ambient pressure $B$ and the phase transition pressure $P_0$. We have plotted the $P-v$ curves (left side of the figures) and the corresponding $v-v_x$ phase diagrams (right side of the figures) for the three types of universes in Fig.~\ref{PV0}, \ref{PV1}, and \ref{PV2}, respectively. Since the situations are similar for all three types of universes, we will discuss the flat universe case ($k=0$) as an example as follows:
\begin{figure}[h]
	\centering
	\includegraphics[width=1\textwidth]{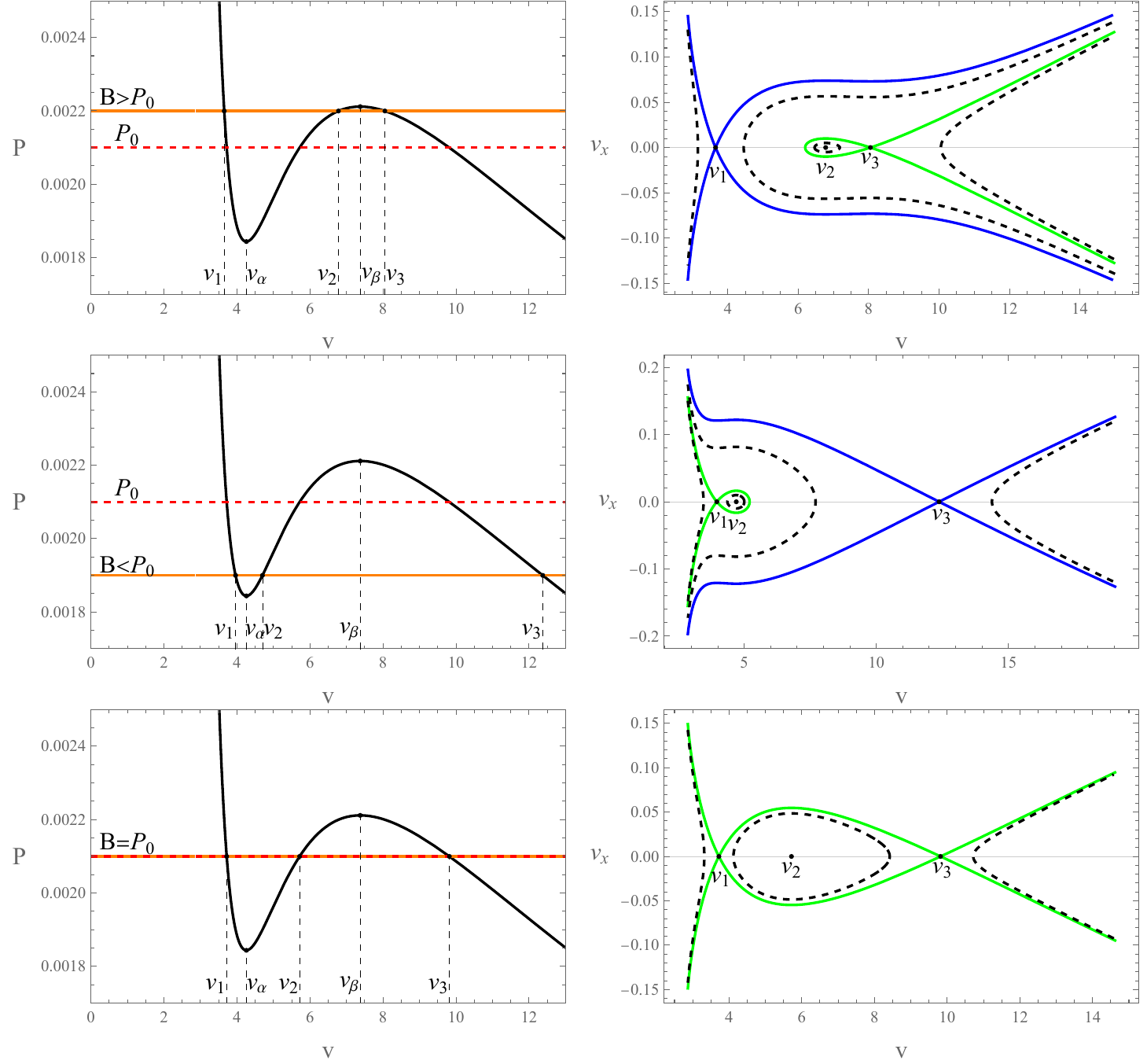}
	\caption{The figure shows the $P-v$ isotherms (left side) and the corresponding unperturbed phase diagrams in the $v-v_x$ phase space (right side) for a flat universe ($k=0$). From top to bottom, they correspond to $P_0<B<P(v_{\beta},T_0)$, $P(v_{\alpha},T_0)<B<P_0$, and $B=P_0$, respectively. We set $\gamma=0.2375$, $T=0.9 T_c\approx0.03559$, and $A=0.2$.}
	\label{PV0}
\end{figure}

\begin{itemize}
	\item Case-1, $P_0<B<P(v_{\beta},T_0)$.\\
	As shown in the upper part of Fig.~\ref{PV0}, there is a homoclinic orbit represented by a green curve connecting the saddle point $v_3$ to itself (with $v_2$ as the center of this orbit), i.e., $v(x)\rightarrow v_3$ as $x\rightarrow\pm\infty$, $v_x(x)\rightarrow 0$ as $x\rightarrow\pm\infty$. The blue curve passing through $v_1$ is neither a homoclinic orbit nor a heteroclinic orbit, while the black dashed lines represent some intermediate state trajectories (this applies to all three cases and will not be repeated later);\\
	\item Case-2, $P(v_{\alpha},T_0)<B<P_0$.\\
	As shown in the middle part of Fig.~\ref{PV0}, similar to case 1, there is a homoclinic orbit represented by a green curve connecting the saddle point $v_1$ to itself (with $v_2$ as the center of this orbit), i.e., $v(x)\rightarrow v_1$ as $x\rightarrow\pm\infty$, $v_x(x)\rightarrow 0$ as $x\rightarrow\pm\infty$;\\
	\item Case-3, $B=P_0$.\\
	As shown in the lower part of Fig.~\ref{PV0}, unlike the previous two cases, there is a heteroclinic orbit represented by a green curve connecting the saddle points $v_1$ and $v_3$ (with $v_2$ as the center of this orbit), i.e., $v(x)\rightarrow v_1$, as $x\rightarrow -\infty$, $v(x)\rightarrow v_3$, as $x\rightarrow +\infty$, $v_x(x)\rightarrow 0$, as $x\rightarrow \pm\infty$.
\end{itemize}

Now, introduce a spatially periodic perturbation, which takes the form
\begin{equation}
	T=T_0+\epsilon\cos qx,
\end{equation}
and then Eq.~(\ref{BB}) becomes

\begin{equation}
	Av''+P(v,T_0)+\frac{\epsilon\cos qx}{v}=B.
\end{equation}
By denoting $v^{\prime}=h$, this second-order differential equation can be rewritten as two first-order differential equations:
\begin{equation}\label{hh}
	\begin{aligned}
		&v^{\prime}=h,\\&h^{\prime}=\frac{B-P(v,T_{0})}{A}-\frac{\epsilon\cos qx}{Av}.
	\end{aligned}
\end{equation}
Repeating the process discussed for Eq.~(\ref{xu}), under spatially periodic perturbations, the forms of $f(z)$ and $g(z)$ are given as
\begin{equation}\label{fzz}
	f(z)=\begin{bmatrix}h\\\frac{B-P(v,T_{0})}{A}\end{bmatrix},
\end{equation}
and
\begin{equation}\label{gzz}
	g(z)=\begin{bmatrix}0\\-\frac{\cos qx}{Av}\end{bmatrix}.
\end{equation}
Denote the solution of Eq.~(\ref{hh}) without perturbations ($\epsilon=0$) as
\begin{equation}\label{hl}
	z_0(x)=\begin{bmatrix}v_0(x)\\h_0(x),\end{bmatrix}.
\end{equation}

The Melnikov function applicable to spatially periodic perturbations is given by
\begin{equation}\label{Mx}
	M\left(x_0\right)=\int_{-\infty}^{+\infty}f^{\mathrm{T}}\left[z_0\left(x-x_0\right)\right]\mathbf{J}g\left[z_0\left(x-x_0\right)\right]dx.
\end{equation}
By substituting Eq.~(\ref{fzz}), (\ref{gzz}), and (\ref{hl}) into Eq.~(\ref{Mx}), we obtain
\begin{equation}
	M(x_0)=-\int_{-\infty}^{+\infty}\frac{h_0(x-x_0)\cos(qx-qx_0)}{Av_0(x-x_0)}dx.
\end{equation}
Expanding the cosine function $cos(qx-qx_0)$, the above expression can be written as
\begin{equation}\label{Mxl}
	M(x_0)=-N\cos(qx_0)-W\sin(qx_0),
\end{equation}
with
\begin{equation}
	\begin{aligned}
		&N=\int_{-\infty}^{+\infty}\frac{h_{0}(x-x_{0})\cos qx}{Av_{0}(x-x_{0})}dx,\\
		&W=\int_{-\infty}^{+\infty}\frac{h_{0}(x-x_{0})\sin qx}{Av_{0}(x-x_{0})}dx.
	\end{aligned}
\end{equation}
Letting $M(x_0)=0$, Eq.~(\ref{Mxl}) simplifies to
\begin{equation}\label{Mxll}
	0=N\cos(qx_0)+W\sin(qx_0),
\end{equation}
and the discussion follows:
\begin{itemize}
	\item $N=W=0$: For any $x_0$, Eq.~(\ref{Mxll}) holds true;
	\item $N=0, W\ne0$: When $x_0=\frac{n \pi}{q}$, Eq.~(\ref{Mxll}) holds true, where $n\in \mathbb{Z}$;
	\item $N\ne0, W=0$: When $x_0=\frac{n \pi+\frac{\pi}{2}}{q}$, Eq.~(\ref{Mxll}) holds true, where $n\in \mathbb{Z}$;
	\item $N\ne0, W\ne0$: When $x_0=\frac{-\arctan \frac{N}{W}}{q}$, Eq.~(\ref{Mxll}) holds true.
\end{itemize}
Therefore, regardless of the perturbation, $M(x_0)$ always has a simple zero, indicating that spatial periodic perturbations will always induce chaotic behavior in the quantum-corrected-AdS black hole fluid system.

\section{Summary}\label{sec5}
In this work, we extend the quantum-corrected black hole to AdS spacetime, briefly discussing its equation of state and critical point behavior in the three types of universes. We then provide a detailed analysis of its chaotic behavior under temporal and spatial perturbations in the extended phase space.

Analysis of the equation of state and critical points of this black hole indicates that its thermodynamic behavior is similar to that of a Van der Waals system.
In all three types of universes, an increase in the quantum correction parameter $\gamma$ leads to a decrease in $T_c$ and $P_c$, and an increase in $v_c$. With $\gamma$ fixed, larger values of $k$ result in higher $T_c$ and $P_c$, and lower $v_c$. An increase in the dust sphere radius $\tilde{r}_0$ has no effect on $T_c$, $P_c$, and $v_c$ in a flat universe but increases $T_c$ and $P_c$ and decreases $v_c$ in a closed universe, while the opposite occurs in an open universe.
Interestingly and importantly, although $T_c$, $P_c$, and $v_c$ all depend on $\gamma$, $\frac{P_cV_c}{T_c}$ does not. At $\tilde{r}_0=0$, the $\frac{P_cV_c}{T_c}$ values for the three universes coincide at $\frac{P_cV_c}{T_c}=\frac{7}{18}$ (because $\tilde{r}_0=0$ causes $k$ to disappear from the metric). As $\tilde{r}_0$ increases, $\frac{P_cV_c}{T_c}$ remains unchanged in the flat universe, increases monotonically in the closed universe, and decreases monotonically in the open universe. Additionally, only in an open universe can $\frac{P_cV_c}{T_c}$ reach the Van der Waals system's value of $\frac{P_cV_c}{T_c}=\frac{3}{8}$, whereas in the other two types of universes, $\frac{P_cV_c}{T_c}$ is always greater than $\frac{3}{8}$.

Starting from the black hole's equation of state, we derived the perturbation Hamiltonian system corresponding to fluid motion in the spinodal region and found that, because $P(v,T)$ is not a linear function of $T$, there are more nonlinear terms in the Hamiltonian compared to other studies. However, these additional nonlinear terms only appear in the higher-order terms of $\epsilon$, which we appropriately ignored. We also presented homoclinic and heteroclinic orbits in phase space and provided information on the occurrence of chaos in the thermodynamic phase space by analyzing the zeros of the Melnikov function.

Under temporal perturbations, when the perturbation amplitude $\delta$ exceeds the critical amplitude $\delta_c$, the evolution of homoclinic orbits corresponding to quantum-corrected-AdS black holes in all three types of universes becomes chaotic and unpredictable, indicating the occurrence of chaotic behavior. When $\tilde{r}_0$ is fixed, the relationship between $\delta_c$ and $\gamma$ shows that $\delta_c$ decreases first and then increases as the quantum correction parameter $\gamma$ increases. This implies that chaos is more difficult to occur when $\gamma$ is relatively small or large. Notably, with $\gamma$ determined to be $0.2375$, chaos is always more likely to occur compared to when $\gamma = 0$ (the Schwarzschild black hole case). As $\tilde{r}_0$ increases from $0$, the $\delta-\gamma$ curves in the three universes separate from coinciding (since $\tilde{r}_0=0$ causes $k$ to disappear from the metric). Specifically, in a flat universe, the $\delta_c-\gamma$ curve remains unchanged (since $k=0$ also causes $\tilde{r}_0$ to disappear from the metric). In regions with smaller $\gamma$ values, the $\delta_c-\gamma$ curve for a closed universe shifts upward, and for an open universe, it shifts downward; in regions with larger $\gamma$ values, the opposite occurs. This means that chaos is most likely to occur in an open universe and least likely in a closed universe when $\gamma$ values are small, while chaos is most likely to occur in a closed universe and least likely in an open universe when $\gamma$ values are large. Moreover, this variation in the $\delta_c-\gamma$ curve indicates that for any two universes, their corresponding $\delta_c-\gamma$ curves will always intersect at some point, which is interesting. For the real scenario (if $\gamma=0.2375$ is correct), the $\delta_c-\tilde{r}_0$ curve with $\gamma$ fixed at $0.2375$ shows that chaos is always more difficult to occur in an open universe. In a flat universe, chaos is more difficult to occur than in a closed universe when $\tilde{r}_0$ is small, but after $\tilde{r}_0$ reaches a certain value, it is always more difficult to occur in a closed universe than in a flat universe. The $\delta_c-\tilde{r}_0$ curve also shows that there is an $\tilde{r}_0$ value that makes chaos most likely to occur in a closed universe. These observations might help us determine the curvature of the universe. 

Under spatial perturbations, we plotted the unperturbed phase diagrams of quantum-corrected-AdS black holes in the $v-v_x$ phase space for the three types of universes. These diagrams are very similar to those of other AdS black holes, with no essential differences. The chaotic behavior is also similar to most AdS black holes, where chaos always occurs regardless of the perturbations.

\backmatter

\bmhead{Acknowledgements}

We are grateful to Shi-Jie Ma and Yu-Hang Feng for their useful suggestions. We also thank the National Natural Science Foundation of China (Grant No.11571342) for supporting us on this work.

\bibliography{sn-bibliography.bib}

\end{document}